\documentstyle[12pt]{article}
\newcommand{\be}{\begin{equation}}
\newcommand{\ee}{\end{equation}}
\newcommand{\bea}{\begin{array}}
\newcommand{\eea}{\end{array}}
\newtheorem{Lemma}{Lemma}
\newtheorem{Proposition}{Proposition}
\newtheorem{Theorem}{Theorem}
\newtheorem{Corollary}{Corollary}
\def\proof{\par{\it Proof}. \ignorespaces} \def\endproof{{\ \vbox{\hrule\hbox{%
\vrule height1.3ex\hskip0.8ex\vrule}\hrule }}\par}
\newenvironment{Proof}{\proof}{\endproof}  

\title{SOLUTION OF THE\\
DISPERSIONLESS HIROTA EQUATIONS}
\author{R. Carroll\thanks{Department of mathematics, University of
Illinois, Urbana, IL 61801 \endgraf {\it E-mail}: \ rcarroll@math.uiuc.edu}
\ \ and Y. Kodama\thanks{Department of Mathematics, Ohio State University,
Columbus, OH 43210 \endgraf {\it E-mail}: \ kodama@math.ohio-state.edu}}

\date{May, 1995}

\begin{document}
\bibliographystyle{plain}
\maketitle

\begin{abstract}
The dispersionless differential Fay identity is shown to be equivalent to a
kernel expansion providing a universal algebraic characterization and
solution of the dispersionless Hirota equations. Some calculations based on
D-bar data of the action are also indicated. \end{abstract}

\section{Introduction}
The dKP hierarchy (= dispersionless limit of the KP hierarchy) and various
reductions thereof (such as dNKdV) play an increasingly important role in
topological field theory and its connections to strings and 2-D gravity
(cf. \cite{aa,ac,af,ca,cj,dp,kk,tb,tc}). There are also connections to
twistor
theory in the spirit of \cite{tb,tc} but we do not pursue this direction
here. We address rather the dispersionless Hirota equations which arise
from the dispersionless differential Fay identity and characterize dKP. We
show that this identity is equivalent to a kernel expansion which generates
the dispersionless Hirota equations in a universal algebraic manner
depending only on the expressions $\lambda = P +
\sum_1^{\infty}U_{n+1}P^{-n}$ or equivalently $P = \lambda -
\sum_1^{\infty}P_{j+1}\lambda^{-j}$. The $P_{j+1}$ play the role of
universal coordinates and this aspect has important consequences in
topological field theory (cf. \cite{af}). The solutions are given in terms
of a residue formula $F_{mn} = Res_P[\lambda^md\lambda^n_{+}]$, which in
fact are shown to be generated
by the universal coordinates $P_{n+1} = F_{1n}/n$. We also show how the
Hamilton-Jacobi and hodograph analysis for dNKdV yields D-bar data for the
action $S$ and this gives a moment type formula for the $F_j$ in terms of
$\bar{\partial}S$. In the dKdV situation this actually yields a direct
formula for the $F_{ij}$ providing an immediate alternative method of
calculation.

\medskip

\section{Background on KP and dKP}
\renewcommand{\theequation}{2.\arabic{equation}}\setcounter{equation}{0}

One can begin with two pseudodifferential operators ($\partial =
\partial/\partial x$),
\be
L = \partial + \sum_1^{\infty}u_{n+1}\partial^{-n};\, \ \,W = 1 + \sum_1^
{\infty}w_n\partial^{-n} \ ,
\label{AA}
\ee
called the Lax operator and gauge operator respectively, where the
generalized Leibnitz rule with $\partial^{-1} \partial = \partial
\partial^{-1} = 1$ applies \be
\partial^if\cdot \ = \sum_{j=0}^{\infty}{i \choose j}(\partial^j f)
\partial^ {i-j} \cdot \ ,
\label{AB}
\ee
for any $i \in {\bf Z}$, and $L = W\partial\,W^{-1}$. The KP hierarchy then
is determined by
the Lax equations ($\partial_n = \partial/\partial t_n$), \be \partial_n L
= [B_n,L] := B_n L - L B_n \ , \label{AC} \ee where $B_n = L^n_{+} $ is the
differential part of $L^n = L^n_{+} + L^n_{-} =
\sum_0^{\infty}\ell_i^n\partial^i + \sum_{-\infty}^
{-1}\ell_i^n\partial^i$. One can also express this via the Sato equation,
\be \partial_n W\,W^{-1} = -L^n_{-} \ ,
\label{AD}
\ee
which is particularly well adapted to the dKP theory. Now define the wave
function
via
\be
\psi = W\,e^{\xi} = w(t,\lambda)e^{\xi};\, \ \,\xi := \sum_1^{\infty}t_n
\lambda^n;
\, \ \,w(t,\lambda) = 1 + \sum_1^{\infty}w_n(t)\lambda^{-n} \ , \label{AE}
\ee where $t_1 = x$. There is also an adjoint wave function $\psi^{*} =
W^{*-1} \exp(-\xi) = w^{*}(t,\lambda)\exp(-\xi),\,\,w^{*}(t,\lambda) = 1 +
\sum_1^ {\infty}w_i^{*}(t)\lambda^{-i}$, and one has equations \be L\psi =
\lambda\psi;\, \ \,\partial_n\psi = B_n\psi;\, \ \,L^{*}\psi^{*} =
\lambda\psi^{*};\, \ \,\partial_n\psi^{*} = -B_n^{*}\psi^{*} \ . \label{AF}
\ee
Note that the KP hierarchy (\ref{AC}) is then given by the compatibility
conditions among these equations with an iso-spectral property. Next one
has the fundamental tau function $\tau(t)$ and vertex operators ${\bf
X},\,\,{\bf X}^{*}$ satisfying
\be
\psi(t,\lambda) = \frac{{\bf X}(\lambda)\tau (t)}{\tau (t)} = \frac{e^{\xi}
G_{-}(\lambda)\tau (t)}{\tau (t)} = \frac{e^{\xi}\tau(t-[\lambda^{-1}])}
{\tau (t)};
\label{AG}
\ee
$$\psi^{*}(t,\lambda) = \frac{{\bf X}^{*}(\lambda)\tau (t)} {\tau (t)} =
\frac{e^{-\xi}
G_{+}(\lambda)\tau (t)}{\tau (t)} = \frac{e^{-\xi}\tau(t+[\lambda^{-1}])}
{\tau (t)} \ ,$$
where $G_{\pm}(\lambda) = \exp(\pm\xi(\tilde{\partial},\lambda^{-1}))$ with
$\tilde{\partial} = (\partial_1,(1/2)\partial_2,(1/3)\partial_3, \cdots)$
and $t\pm[\lambda^{-1}]= (t_1\pm \lambda^{-1},t_2 \pm (1/2) \lambda^{-2},
\cdots)$.
One writes also
\be
e^{\xi} := \exp \left({\sum_1^{\infty}t_n\lambda^n}\right) = \sum_0^
{\infty}\chi_j(t_1, t_2, \cdots ,t_j) \lambda^j \ , \label{AH} \ee where
the $\chi_j$ are the elementary Schur polynomials, which arise in many
important formulas (cf. below). \\[3mm]\indent We mention now the famous
bilinear identity which generates the entire KP hierarchy. This has the
form \be
\oint_{\infty}\psi(t,\lambda)\psi^{*}(t',\lambda)d\lambda = 0 \label{AI}
\ee where $\oint_{\infty}(\cdot)d\lambda$ is the residue integral about
$\infty$, which we also denote $Res_{\lambda}[(\cdot)d\lambda]$. Using
(\ref{AG}) this can also be written in terms of tau functions as \be
\oint_{\infty}\tau(t-[\lambda^{-1}])\tau(t'+[\lambda^{-1}])
e^{\xi(t,\lambda)-\xi(t',\lambda)}d\lambda = 0 \ . \label{AJ} \ee This
leads to the characterization of the tau function in bilinear form
expressed via ($t\to t-y,\,\,t'\to t+y$) \be
\left(\sum_0^{\infty}\chi_n(-2y)\chi_{n+1}(\tilde D)e^{\sum_1^{\infty} y_i
D_i}\right)\tau\,\cdot\,\tau = 0 \ , \label{AK} \ee where $D_i$ is the
Hirota derivative defined as $D^m_j a\,\cdot\,b = (\partial^m/\partial
s_j^m) a(t_j+s_j)b(t_j-s_j)|_{s=0}$ and $\tilde D = (D_1,(1/2)
D_2,(1/3)D_3,\cdots)$. In particular, we have from the coefficients of
$y_n$ in (\ref{AK}), \be \label{hirota}
D_1D_n\tau \cdot \tau = 2 \chi_{n+1} (\tilde D) \tau \cdot \tau \ , \ee
which are called the Hirota bilinear equations. Such calculations with
vertex operator
equations and residues, in the context of finite zone situations where the
tau function is intimately related to theta functions, also led
historically to the Fay trisecant identity, which can be expressed
generally as the Fay identity via (cf.
\cite{aa,ce} - c.p. means cyclic permutations) \be
\sum_{c.p.}(s_0-s_1)(s_2-s_3)\tau(t+[s_0]+[s_1])\tau(t+[s_2]+[s_3]) = 0 \ .
\label{ZA}
\ee
This can be also derived from the bilinear identity (\ref{AJ}). Differentiat
ing this in $s_0$, then setting $s_0 = s_3 = 0$, then dividing by $s_1
s_2$, and finally shifting $t\to t-[s_2]$, leads to the differential Fay
identity,
\begin{eqnarray}
\nonumber
& &\tau(t)\partial\tau(t+[s_1]-[s_2]) - \tau(t+[s_1] -[s_2])\partial
\tau(t) \\ & &= (s_1^{-1}-s_2^{-1}) \left[\tau(t+[s_1]-[s_2]) \tau(t) -
\tau(t+[s_1])\tau(t-[s_2])\right] \ . \label{AL} \end{eqnarray} The Hirota
equations
(\ref{hirota}) can be also derived from (\ref{AL}) by taking the limit $s_1
\to s_2$. The identity (\ref{AL}) will play an important role later.
\\[3mm]\indent
Now for the dispersionless theory (dKP) one can think of fast and slow
variables, etc., or averaging procedures, but simply one takes
$t_n\to\epsilon t_n = T_n\,\,(t_1 = x\to \epsilon x = X)$ in the KP
equation $u_t = (1/4)u_{xxx} + 3uu_x + (3/4)\partial^{-1}u_{yy},\,\,
(y=t_2,\,\,t=t_3)$, with $\partial_n\to \epsilon\partial/\partial T_n$ and
$u(t_n)\to U(T_n)$ to obtain $\partial_T U = 3UU_X + (3/4)\partial^
{-1}U_{YY}$ when $\epsilon\to 0\,\,(\partial =\partial/\partial X$ now).
Thus the dispersion term $u_{xxx}$ is removed. In terms of hierarchies we
write \be
L_{\epsilon} = \epsilon\partial + \sum_1^{\infty}u_{n+1}(T/\epsilon)
(\epsilon\partial)^{-n} \ ,
\label{AM}
\ee
and think of $u_n( T/\epsilon)= U_n(T) + O(\epsilon)$, etc. One takes then
a WKB form for the wave function with the action $S$ {\cite{ka}}, \be \psi
= \exp \left[\frac{1}{\epsilon}S(T,\lambda) \right]. \label{AN} \ee
Replacing now $\partial_n$ by $\epsilon\partial_n$, where $\partial_n =
\partial/\partial T_n$ now, we define $P := \partial S = S_X$. Then
$\epsilon^i\partial^i\psi\to P^i\psi$ as $\epsilon\to 0$ and the equation
$L\psi = \lambda\psi$ becomes
\be
\lambda = P + \sum_1^{\infty}U_{n+1}P^{-n};\, \ \,P = \lambda -
\sum_1^{\infty}P_{i+1}\lambda^{-i} \ , \label{AO} \ee where the second
equation is simply the inversion of the first. We also note from
$\partial_n\psi =
B_n\psi = \sum_0^nb_{nm}(\epsilon\partial)^m\psi$ that one obtains
$\partial_n S = {\cal B}_n(P) = \lambda^n_{+}$ where the subscript (+)
refers now to powers of $P$ (note $\epsilon\partial_n\psi/\psi \to
\partial_n S$). Thus $B_n = L^n_{+}\to {\cal B}_n(P) = \lambda^n_{+} =
\sum_0^nb_{nm}P^m$ and the KP hierarchy goes to \be \partial_n P = \partial
{\cal B}_n \ ,
\label{YC}
\ee
which is the dKP hierarchy
(note $\partial_n S = {\cal B}_n\Rightarrow \partial_n P = \partial{\cal
B}_n$).
The action $S$ in (\ref{AN}) can be computed from (\ref{AG}) in the limit
$\epsilon \to 0$ as
\be
\label{action}
S = \sum_{1}^{\infty} T_n \lambda^n - \sum_{1}^{\infty} {\partial_mF \over
m} \lambda^{-m},
\ee
where the function $F=F(T)$ (free energy) is defined by {\cite{tb}} \be
\label{tau}
\tau = \exp \left[ {1 \over \epsilon^2} F(T) \right]. \ee The formula
(\ref{action}) then solves the dKP hierarchy (\ref{YC}), i.e. $P={\cal B}_1
= \partial S$ and
\be
\label{B}
{\cal B}_n = \partial_n S =
\lambda^n - \sum_{1}^{\infty} {F_{nm} \over m} \lambda^{-m} \ , \ee where
$F_{nm} = \partial_n\partial_m F$ which play an important role in the
theory of dKP.
\\[3mm]\indent
Now following \cite{tc} we write the differential Fay identity (\ref{AL})
with $\epsilon\partial_n$ replacing $\partial_n$ etc. in the form
\begin{eqnarray}
\label{AR}
\nonumber
& &\frac{\tau(T-\epsilon[\mu^{-1}]-\epsilon[\lambda^{-1}])\tau(T)}
{\tau(T-\epsilon[\mu^{-1}])\tau(T-\epsilon[\lambda^{-1}])} \\ & & = 1 +
\frac{\epsilon\partial}{\mu-\lambda}\left[\log\left( \tau(T-
\epsilon[\mu^{-1}])\right)
- \log\left(\tau(T-\epsilon[\lambda^{-1}])\right)\right] \end{eqnarray} (in
(\ref{AL}) take $t\to t-[s_1],\,\,s_1=\mu^{-1},\,\,s_2=\lambda^{-1}$ and
insert $\epsilon$ at the appropriate places; note $T$ is used in
(\ref{AR})). One notes from
(\ref{AH}) that $\exp(-\xi(\tilde{\partial},\lambda^{-1})) = \sum_1^
{\infty}\chi_j(\tilde{\partial})\lambda^{-j}$ so taking logarithms in
(\ref{AR}) and using (\ref{tau}) yield \begin{eqnarray} \label{AS}
\nonumber
& &{1 \over \epsilon^2} \sum_{m,n=1}^{\infty}\mu^{-m} \lambda^
{-n}\chi_n(-\epsilon\tilde{\partial}) \chi_m(-\epsilon\tilde{\partial})F \\
& &= \log\left[1+ {1 \over \epsilon} \sum_1^{\infty}\frac{\mu^{-n}-
\lambda^{-n}}{\mu-\lambda}
\chi_n(-\epsilon\tilde{\partial})\partial_XF \right] \ . \end{eqnarray} In
passing this to limits only the second order derivatives survive, and one
gets the dispersionless differential Fay identity (note $\chi_n
(-\epsilon\tilde{\partial})$ contributes here only $-\epsilon\partial_n/n$)
\be \sum_{m,n=1}^{\infty}\mu^{-m}\lambda^{-n}\frac{F_{mn}}{mn} = \log
\left(1- \sum_1^{\infty}\frac{\mu^{-n}-\lambda^{-n}}{\mu-\lambda}
\frac{F_{1n}}{n} \right) \ .
\label{AT}
\ee
Although (\ref{AT}) only uses a subset of the Pl\"ucker relations defining
the KP hierarchy it was shown in \cite{tc} that this subset is sufficient
to determine KP; hence (\ref{AT}) characterizes the function $F$ for dKP.
Following \cite{cb}, we now derive a dispersionless limit of the Hirota
bilinear equations (\ref{hirota}), which we call the dispersionless Hirota
equations. We first note from (\ref{action}) and (\ref{AO}) that $F_{1n} =
nP_{n+1}$
so
\be
\sum_1^{\infty}\lambda^{-n}\frac{F_{1n}}{n} = \sum_1^{\infty}P_{n+1}
\lambda^{-n} = \lambda - P(\lambda) \ .
\label{AU}
\ee
Consequently the right side of (\ref{AT}) becomes $\log[\frac{P(\mu) -
P(\lambda)}{\mu-\lambda}]$ and for $\mu\to \lambda$ with $\dot{P}:=
\partial_{\lambda}P$ we have
\be
\log\dot{P}(\lambda) = \sum_{m,n=1}^{\infty}\lambda^{-m-n}\frac{F_{mn}}
{mn} = \sum_{j=1}^{\infty} \left(\sum_{n+m=j} {F_{mn} \over mn} \right)
\lambda^{-j} \ .
\label{AV}
\ee
Then using the elementary Schur polynomial defined in (\ref{AH}) and
(\ref{AO}), we obtain
\begin{eqnarray}
\nonumber
\dot{P}(\lambda) &=& \sum_0^{\infty} \chi_j(Z_2, \cdots,Z_j) \lambda^{-j}
\\ &=&1 + \sum_1^{\infty}jP_{j+1}\lambda^{-j-1} = 1 + \sum_1^{\infty}F_{1j}
\lambda^{-j-1} \ , \label{AW}
\end{eqnarray}
where $Z_i, \, i \ge 2$ are defined by
\be
\label{SV}
Z_i = \sum_{m+n=i} {F_{mn} \over mn} \ . \ee Note that $Z_1=0$ is assumed
for the polynomials $\chi_j$. Thus we obtain the dispersionless Hirota
equations, \be \label{F}
F_{1j} = \chi_{j+1}(Z_2, \cdots,Z_{j+1}) \ . \ee These can
be also derived directly from (\ref{hirota}) with (\ref{tau}) in the limit
$\epsilon \to 0$ or by expanding (\ref{AV}) in powers of $\lambda^{-n}$.
We list here a few entries from such an expansion (cf. \cite{cb}):
\begin{eqnarray} \lambda^{-4}&:& \frac{1}{2}F^2_{11} - \frac{1}{3}F_{13} +
\frac{1}{4}
F_{22} = 0;\label{AX}\\
\lambda^{-5}&:& F_{11}F_{12} - \frac{1}{2}F_{14} + \frac{1}{3}F_{23} = 0;
\nonumber\\
\lambda^{-6}&:& \frac{1}{3}F^3_{11} - \frac{1}{2}F^2_{12} - F_{11}F_{13} +
\frac{3}{5}F_{15} - \frac{1}{9}F_{33} - \frac{1}{4}F_{24} = 0; \nonumber\\
\lambda^{-7}&:& F^2_{11}F_{12} - F_{12}F_{13}-F_{11}F_{14} +
\frac{2}{3}F_{16} - \frac{1}{6}F_{34} - \frac{1}{5}F_{25} = 0; \nonumber\\
\lambda^{-8}&:& \frac{1}{4}F^4_{11} - F_{11}F_{12}^2 - F^2_{11}F_{13} +
\frac{1}{2}F^2_{13} + F_{12}F_{14}+\nonumber\\ & & + F_{11}F_{15} -
\frac{5}{7}F_{17} + \frac{1}{16}F_{44} + \frac{2}{15} F_{35} +
\frac{1}{6}F_{26} = 0;\nonumber\\ \lambda^{-9}&:& F^3_{11}F_{12} -
\frac{1}{3}F^3_{12} - 2F_{11}F_{12}F_{13} -F^2_{11}F_{14} + F_{13}F_{14}
\nonumber\\
& & + F_{12}F_{15} + F_{11}F_{16} - \frac{3}{4}F_{18} + \frac{1}{10}F_{45}
+ \frac{1}{9}F_{36} + \frac{1}{7}F_{27} = 0;\nonumber\\ \lambda^{-10}&:&
\frac{1}{5}F^5_{11}-\frac{3}{2}F^2_{11}F^2_{12} -F^3_{11}F_{13} +
F^2_{12}F_{13} + F_{11}F^2_{13}\nonumber\\ & & +2F_{11}F_{12}F_{14} -
\frac{1}{2}F^2_{14} + F^2_{11}F_{15} - F_{13}F_{15} -F_{12}F_{16}
-F_{11}F_{17}\nonumber\\ & & +\frac{7}{9}F_{19} -\frac{1}{25}F_{55}
-\frac{1}{12}F_{46} - \frac{2}{21} F_{37} - \frac{1}{8}F_{28} = 0 \ .
\nonumber \end{eqnarray} These equations are discussed in various ways
below and we will also show the equivalence of the dispersionless Fay
differential identity with another formula of a Cauchy kernel in Section 3.
Note here that for $U = F_{11}$ the first equation in (\ref{AX}) is a dKP
equation $U_T = 3UU_X + (3/4)\partial^{-1} U_{YY}$ and other equations in
the hierarchy are similarly generated.

\medskip

It is also interesting to note that the dispersionless Hirota equations
(\ref{F}) or (\ref{AX}) can be regarded as algebraic equations for
``symbols" $F_{mn}$, which are defined via (\ref{B}), i.e. \be
{\cal B}_n := \lambda^n_+= \lambda^n - \sum_1^{\infty}\frac{F_{nm}}{m}
\lambda^{-m}\ .
\label{AY}
\ee
\begin{Lemma}
The symbols satisfy
\be
F_{nm} = F_{mn} = Res_P[\lambda^m d \lambda^n_+] \ . \label{AZ} \ee \end{Lemma}
\begin{Proof}
One need simply observe that
\begin{eqnarray}
\nonumber
F_{nm} &=&- Res_{\lambda}[{\cal B}_nd\lambda^m] = - Res_P[{\cal
B}_nd\lambda^m] \\ \nonumber
&=& - Res_P[\lambda^n_{+}d\lambda^m_{-}] =- Res_P[\lambda^nd\lambda^m_{-}]
\\ &=& Res_P[\lambda^nd\lambda^m_{+}] = Res_P[\lambda^md\lambda^n_{+}] \ =
\ F_{mn}.
\label{BA}
\end{eqnarray}
Here we have used $\lambda^m_{-} = \lambda^m - \lambda^m_{+}$ and
$Res[d(ab)] = 0
= Res[da\,b + a\,db]$ for pseudo-differential or formal Laurent expansions
$a$ and $b$.
\end{Proof}

\medskip
\noindent
Thus we have:
\begin{Theorem} For $\lambda,\,\,P$ given algebraically as in (\ref{AO}),
with no a priori connection to dKP, and for ${\cal B}_n$ defined as in the
last equation of (\ref{AY}) via a formal collection of the symbols with two
indices $F_{mn}$, it follows that the dispersionless Hirota equations
(\ref{F}) or (\ref{AX}) are nothing but the polynomial identities among
$F_{mn}$. \end{Theorem}

In the next section we give a direct proof of this fact, that is, the
$F_{mn}$ defined in (\ref{AY}) satisfy the dispersionless
Hirota equations, which we designate by (\ref{F}) in what follows. \medskip

Now one very natural way of developing dKP begins with (\ref{AO}) and
(\ref{YC}) since
eventually the $P_{j+1}$ can serve as universal coordinates (cf. here
\cite{af} for a discussion of this in connection with topological field
theory = TFT). This point of view is also natural in terms of developing a
Hamilton-Jacobi theory involving ideas from the hodograph $-$ Riemann
invariant approach (cf. \cite{ca,gf,ka,kc} and (\ref{YE}) below) and in
connecting
NKdV ideas to TFT, strings, and quantum gravity (cf. \cite{cj} for a survey
of this).
It is natural here to work with $Q_n := (1/n){\cal B}_n$ and note that
$\partial_n S = {\cal B}_n$ corresponds to $\partial_n P = \partial {\cal
B}_n = n\partial Q_n$.
In this connection one often uses different time variables, say $T'_n =
nT_n$, so that $\partial'_nP = \partial Q_n$, and $G_{mn} = F_{mn}/mn$ is
used in place of $F_{mn}$. Here however we will retain the $T_n$ notation
with $\partial_n S = nQ_n$ and $\partial_n P = n\partial Q_n$ since one
will be connecting a number of formulas to standard KP notation. Now given
(\ref{AO}) and (\ref{YC}) the equation
$\partial_n P = n\partial Q_n$ corresponds to Benney's moment equations and
is equivalent to a system of Hamiltonian equations defining the dKP
hierarchy (cf. \cite{ca,kc}
and remarks after (\ref{YC}); the Hamilton-Jacobi equations are $\partial_n
S = nQ_n$ with Hamiltonians $nQ_n(X, P=\partial S)$).

\medskip

We now have an important formula for the functions $Q_n$:
\begin{Proposition} \cite{kc} The generating function of $\partial_P
Q_n(\lambda)$ is given by \be \frac{1}{P(\mu) - P(\lambda)} =
\sum_1^{\infty}\partial_P Q_n(\lambda) \mu^{-n} \ . \label{BF}
\ee
\end{Proposition}
\begin{Proof}
Multiplying (\ref{AO})
by $\lambda^{n-1}\partial_P\lambda$, we have \be \lambda^n\partial_P\lambda
= P(\lambda)\lambda^{n-1}\partial_P\lambda + P_2\lambda^{n-2}\partial_P
\lambda + \cdots \ . \label{BG} \ee
Taking the polynomial part leads to a recurrence relation \be
\partial_PQ_{n+1}(\lambda) = P(\lambda)\partial_P Q_n (\lambda)+
P_2\partial_P Q_{n-1}(\lambda) + \cdots + P_n\partial_P Q_1(\lambda) \ .
\label{BH}
\ee
Then noting $\partial_PQ_1=1, Q_0=1$, and summing up (\ref{BH}) as follows,
we obtain
\begin{eqnarray}
\label{sum}
\nonumber
& &\sum_{n=0}^{\infty} \left( \partial_P Q_{n+1}(\lambda) - P(\lambda)
\partial_P
Q_n(\lambda) - \sum_{i+j=n} P_{j+1} \partial_P Q_i(\lambda) \right)\mu^{-n}
\\ & & = \left( \mu - P(\lambda) - \sum_1^{\infty} P_{j+1}\mu^{-j} \right)
\sum_1^{\infty} \partial_P Q_i(\lambda) \mu^{-i} = 1 \ , \end{eqnarray}
which is just (\ref{BF}).
\end{Proof}

\medskip

This is a very important kernel formula which will come up in various ways
in what follows. In particular we note \be \oint_{\infty} {\mu^n \over
P(\mu) - P(\lambda)} d\mu = \partial_P Q_{n+1}(\lambda) \ ,
\label{canonicalT}
\ee
which gives a key formula in the Hamilton-Jacobi method for the dKP
\cite{kc}. Thus it represents a Cauchy kernel and has a version on Riemann
surfaces related to the prime form (cf. \cite{kd}). In fact the kernel is a
dispersionless limit of the Fay prime form. Also note here that the
function $P(\lambda)$ alone provides all the information necessary for the
dKP theory. This will be discussed further in the next section.

\section{Solutions and variations}
\renewcommand{\theequation}{3.\arabic{equation}}\setcounter{equation}{0}

We have already discussed solutions of the dispersionless Hirota equations
(\ref{F}) briefly in Theorem 1,
but we go now to some different points of view. First let us prove:
\begin{Theorem}
The kernel formula (\ref{BF}) is
equivalent to the dispersionless differential Fay identity (\ref{AT}).
\end{Theorem}
This will follow from the following lemma. \begin{Lemma} Let
$\chi_n(Q_1,\cdots,Q_n)$
denote the elementary Schur polynomials defined as in (\ref{AH}). Then \be
\label{ShP}
\partial_P Q_n = \chi_{n-1}(Q_1,\cdots,Q_{n-1}) \ . \ee \end{Lemma}
\begin{Proof}
We integrate the kernel formula
(\ref{BF}) with respect to $P(k)$, and normalize at $\lambda = \infty$ to
obtain \be
\frac{1}{P(\lambda) - P(k)} = \frac{1}{\lambda}\exp \left({\sum_1^{\infty}
Q_n(k)\lambda^{-n}}\right) = \sum_1^{\infty}\chi_{m-1}(Q_1,\cdots,Q_{m-1})
\lambda^{-m} \ .
\label{BI}
\ee
Equating this to (\ref{BF}) gives the result. \end{Proof} \medskip

{\it Proof of Theorem 2}. $ \ \ $ Using (\ref{B}), the left hand side of
(\ref{AT}) can be written as
\begin{eqnarray}
\label{BJ}
\nonumber
LHS &=& \sum_1^{\infty}\frac{1}{n\lambda^n}
\left[{\sum_1^{\infty}\frac{F_{mn}} {m\mu^m}}\right] =
\sum_1^{\infty}\frac{1}{n\lambda^n} [\mu^n - {\cal B}_n(\mu)] \\ &=&
\sum_1^{\infty}\frac{1}{n} \left({\frac{\mu}{\lambda} }\right)^n -
\sum_1^{\infty} \frac{1}{n\lambda^n}{\cal B}_n(\mu) = -\log
\left({1-\frac{\mu}{\lambda} }\right) -
\sum_1^{\infty}\frac{Q_n(\mu)}{\lambda^n} \ . \end{eqnarray}
 On the other hand the right hand side of (\ref{AT}) is calculated as
indicated in (\ref{AU}) and remarks thereafter.  Thus
\begin{eqnarray} \label{BK} \nonumber
RHS &=&
\log \left[{1 - \frac{1}{\lambda - \mu}\sum_1^{\infty}
\left({\frac{1}{\lambda^n} -
\frac{1}{\mu^n}}\right)\frac{F_{1n}}{n}}\right] \\
\nonumber &=& -\log(\lambda - \mu) +
\log \left[{ \left({\lambda - \sum_1^{\infty}\frac{F_{1n}}
{n\lambda^n}} \right) -
\left({\mu - \sum_1^{\infty}\frac{F_{1n}}{n\mu^n}}\right)}
\right] \\ &=& -\log(\lambda - \mu)
+ \log [P(\lambda) - P(\mu)] \ .
\end{eqnarray}
This implies
\be
\log[P(\lambda) - P(\mu)] = \log\lambda -\sum_1^{\infty}{Q_n(\mu)}
{\lambda^{-n}} \ ,
\label{BL}
\ee
which leads to (\ref{BI}). Then Lemma 2 implies the assertion.

\medskip
\noindent
Theorem 2 implies that the dispersionless Hirota equations (\ref{F}) can be
derived from the kernel formula (\ref{BF}) which is a direct consequence of
the definition of $F_{mn}$ in (\ref{AY}). Thus Theorem 2 gives a direct
proof of Theorem 1.

\medskip

We will now express the $\chi_n(Q_1,\cdots,Q_n)$ as polynomials in $Q_1 =
P$ with the coefficients given by polynomials of $P_{j+1}$. First we have:
\begin{Lemma}
One can write
\be
\chi_n = det
\left[
\bea{ccccccc}
P & -1 & 0 & 0 & 0 & \cdots & 0\\
P_2 & P & -1 & 0 & 0 & \cdots & 0\\
P_3 & P_2 & P & -1 & 0 & \cdots & 0\\
\vdots & \vdots & \vdots & \vdots & \vdots & \ddots & \vdots\\ P_n &
P_{n-1} & \cdots & P_4 & P_3 & P_2 & P \eea \right] = \partial_P Q_{n+1} \
{}.
\label{BM}
\ee
\end{Lemma}
\begin{Proof}
In terms of $\chi_n(Q_1,\cdots,Q_n) = \partial_P Q_{n+1}$ the recursion
relation (\ref{BH}) can be written in the form $\chi_n = P\chi_{n-1} +
P_2\chi_{n-2} + \cdots + P_n\chi_0$. It is easy to derive the determinant
expression from this form. \end{Proof}

\medskip

This leads to a rather evident fact, which we express as a proposition
because of its importance. Thus: \begin{Proposition} The $F_{mn}$ can be
expressed
as polynomials in $P_{j+1} = F_{1j}/j$.
\end{Proposition}
\begin{Proof}
$\partial_{\lambda}Q_{n} = \chi_{n-1}(Q_1;P_2,\cdots,P_{n-1})
\partial_{\lambda}Q_1 \,\,(Q_1 = P)$. Put this together with $Q_n= {\cal
B}_n/n$ in (\ref{B}) and $P $ in (\ref{AO}) to arrive at the conclusion
(cf. also Remark 1). \end{Proof} \begin{Corollary}
The dispersionless Hirota
equations can be solved totally algebraically via $F_{mn} =
\Phi_{mn}(P_2,P_3,\cdots,P_{m+n})$ where $\Phi_{mn}$ is a polynomial in the
$P_{j+1}$. Thus the $F_{1n} = nP_{n+1}$ are generating elements for the
$F_{mn}$.
\end{Corollary}
This is of course evident from the dispersionless differential Fay identity
(\ref{AT}). The point here however is to give explicit formulas of $F_{mn}$
in terms of the elementary Schur polynomials $\chi_n$ in (\ref{BM}).

\medskip

\noindent
{\bf Remark 1} $\,\,$ One can also arrive at this polynomial dependence via
the residue formulas of Lemma 1. As an adjunct to the proof of Proposition
2 we note now the following explicit calculations. Recall that $Q_1 = P =
\lambda - \sum_1^{\infty}P_{j+1}\lambda^{-j}$ and generally $Q_n =
\lambda^n/n - \sum_1^{\infty}G_{jn}\lambda^{-j} \ (G_{jn} :=F_{jn}/jn)$.
Thus from $Q_2 = \lambda^2_{+}/2 = P^2/2 + P_2$ we get \be
\frac{1}{2}\left(\lambda -
\sum_1^{\infty}\frac{P_{j+1}}{\lambda^j}\right)^2 + P_2
=\frac{\lambda^2}{2} -\sum_1^{\infty}\frac{G_{2j}}{\lambda^j} \ .
\label{YY}
\ee
Writing this out yields
\be
G_{2m} = P_{2+m} -\frac{1}{2}\sum_{j+k=m}P_{j+1}P_{k+1} \ . \label{YB} \ee
This process can be continued with $Q_3 = \lambda^3_{+}/3 = P^3/3 + P_2P +
P_3$, etc.

\medskip

It is also interesting to note the following relations: \begin{Proposition}
For any $n \ge 2$,
\be
\label{SchurP}
\chi_n(-Q_1,\cdots,-Q_n) =
-P_n .
\ee
\end{Proposition}
\begin{Proof}
obtain (cf. (\ref{AH})) \be
\left(\sum_0^{\infty}\frac{\chi_n(Q)}{\lambda^n}\right)
\left(\sum_0^{\infty}\frac{\chi_m(-Q)}{\lambda^m}\right) =
\sum_0^{\infty}\frac{1}{\lambda^p}\sum_0^p\chi_{p-k}(Q)\chi_k(-Q) = 1 \ .
\label{BN}
\ee
Hence $\sum_0^p\chi_{p-k}(Q)\chi_k(-Q) = 0$ for $p\geq 1$ which implies
$\chi_k(-Q) = -P_k$ for $k\geq 2$ from the recursion relation (\ref{BH}).
\end{Proof}

\medskip

\noindent
Note from $P_{n+1}=F_{1n}/n$ that the equations (\ref{SchurP}) give another
representation of the dispersionless Hirota equations (\ref{F}).

\medskip

\noindent
{\bf Remark 2} $\,\,$ Formulas such as (\ref{BM}) and the statement in
Proposition 2 indicate that in fact dKP theory can be characterized using
only elementary Schur polynomials since these provide all the information
necessary for the kernel (\ref{BF}) or equivalently for the dispersionless
differential Fay identity. This amounts
also to observing that in the passage from KP to dKP only certain Schur
polynomials survive the limiting process $\epsilon\to 0$. Such terms
involve second derivatives of $F$ and these may be characterized in terms
of Young diagrams with only vertical or horizontal boxes. This is also
related to the explicit form of the hodograph transformation where one
needs only $\partial_P Q_n = \chi_{n-1}(Q_1,\cdots,Q_{n-1})$ and the
$P_{j+1}$ in the expansion of $P$ (cf. here (\ref{YE})).

\section{Connections to D-bar}
\renewcommand{\theequation}{4.\arabic{equation}}\setcounter{equation}{0}

It was shown in \cite{ca,cb,gb,za} how inverse scattering information is
connected to the dispersionless theory for KdV and some other situations
(Benney equations and vector nonlinear Schr\"odinger equations for
example). We will see here that although such connections seem generally
not to be expected, nevertheless, one can isolate D-bar data for $S$ and
$P$ in the dispersionless NKdV situations leading to an expression for the
generating elements $F_{1n}=nP_{n+1}$ which can be useful in computation.
The technique also indicates another role for the Cauchy type kernel
$1/(P(\lambda) -P(\mu))$ in (\ref{BF}).
Thus as background
consider KdV and the scattering problem in the standard form (cf.
\cite{ce}) \be u_t = u''' -6uu';\, \ \ \, \psi'' - u\psi = -k^2\psi \ .
\label{CA} \ee
Let $\psi_{\pm}(k,x) \sim \exp(\pm ikx)$ as $x\to \pm\infty$ be Jost
solutions with scattering data, $T$ and $R$, determined via \be \label{TR}
T(k)\psi_{-}(k,x) = R(k)\psi_{+}(k,x) + \psi_{+}(-k,x) \ . \ee Setting
$\psi_{-} = \exp(-ikx + \phi(k,x))$ we have from (\ref{CA}) \be v := -
\partial \log (\psi_-) =ik - \phi'; \, \, \ \phi'' - 2ik\phi' + \phi'^2 = u
\ .
\label{CB}
\ee
One looks for expansions,
\be
\phi'= \sum_1^{\infty}{\phi_n \over (ik)^n};\, \ \ \, v= ik +
\sum_1^{\infty}{v_n \over (ik)^n} \ ,
\label{CC}
\ee
entailing $\phi_n = -v_n$. Here for example one can assume
$\int_{-\infty}^{\infty}(1 + x^2)|u|dx < \infty$ or $u\in {\cal S}$ =
Schwartz space for
convenience so that all of the inverse scattering machinery applies. Then
$T(k)$ will be meromorphic for $Im(k) > 0$ with (possibly) a finite number
of simple poles at $k_j = i\beta_j\,\,(\beta_j > 0),\,\, |R(k)|$ is small
for large $|k|,\,\,k\in {\bf R}$, and $
\log(T)=\sum_0^{\infty}c_{2n+1}/k^{2n+1}$ where $c_n$ can be written in
terms of the $\beta_j$ and the normalization of the wave function
$\psi_{\pm}$. For $x\to\infty,\,\,Im(k) > 0,\,\,\psi_{-} \exp(ikx) \to 1/T$
from
(1.2), and taking logarithms $\phi(k,\infty) = -\log(T)$ which implies \be
-\log(T) = \sum_1^{\infty}\frac{1}{(ik)^n}\int_{-\infty}^{\infty} \phi_n dx
\label{CD}
\ee
leading to $\int_{-\infty}^{\infty} \phi_{2m} dx = 0$ with $ i^{2m}c_{2m+1}
= \int_{-\infty}^{\infty}\phi_{2m+1} dx$. Hence the scattering information
($\sim c_{2n+1}$, arising from $x$-asymptotics, given via $R$ and $T$) is
related to the $k$-asymptotics $\phi_n$ of the wave function $\psi_{-}$.
This kind of connection between space asymptotics (or spectral data - which
is generated by these asymptotic conditions and forced by boundary
conditions on $u$ here) and the spectral asymptotics of the wave function
is generally more complicated and we refer to a formula in the Appendix to
\cite{ca} describing the Davey-Stewartson (DS) situation \be -2\pi
i\chi_{n+1} = \int_{\Omega}\int\zeta^n\bar{\partial}_{\zeta} \chi \ d\zeta
\wedge d\bar{\zeta} \ ,
\label{CE}
\ee
where $\bar{\partial}_{\zeta}\chi$ is D-bar data and the wave function
$\psi$ has the form $\psi = \chi \exp(\Xi)$ with $\chi = 1 +
\sum_1^{\infty}\chi_j\lambda^{-j}$ for $|\lambda|$ large (this is in a
matrix form). The potentials in (\ref{CE}) occur in $\chi_1$. The D-bar
data, or departure from analyticity of $\chi$, corresponds to spectral data
in a sometimes complicated way and we refer to
\cite{ab,bd,ca,cf,cl,fb,gh,kg,lb,ma} for more on this. \\[3mm]\indent Now
one expects some connections between inverse scattering for KdV and the
dKdV theory since $\epsilon x = X$ means e.g. $x\to \pm \infty\sim
\epsilon\to 0$ for $X$ fixed. Now from (\ref{TR}), we have \be
{1 \over T }= \frac{1}{2ik}W(\psi_{+},\psi_{-}) := \frac{1}{2ik}
(\psi'_{+}\psi_{-} - \psi_{+}\psi'_{-}) \ . \label{CF} \ee Given
$\psi_{+}\sim\psi = \exp(S/\epsilon)$ and $\psi_{-}\sim \psi^{*}=
\exp(-S/\epsilon)$ (cf. \cite{ca,cb}) one sees that $W(\psi_+,\psi_-) \to
2\partial S/\partial X = 2P$, and from (\ref{AO}) with $\lambda=ik$, \be
\label{T}
{1 \over T} = {P \over ik} = 1 - \sum_1^{\infty} {P_{n+1} \over (ik)^{n+1}}
\ . \ee
Now $\log|T|$ and $\arg(R/T)$ have natural roles as action angle variables
( cf. \cite{cb}) and from
$R = W(\psi_{-}(k,x),\psi_{+}(-k,x))/ W(\psi_{+}, \psi_{-})$ one obtains \be
R = -\left({P(-k) + P(k) \over 2P(k)}\right) \exp\left({\frac{S(k) -
S(-k)}{\epsilon}} \right) \ .
\label{CG}
\ee
(We have suppressed the slow variables $T_n$ ). Thus the scattering data,
which are considered to be averaged over the fast variables, now depend on
the slow variables (in particular, see (4.5) and (4.8)). This corresponds
to a dynamics of
the Riemann surface determined by the scattering problem (the Whitham
averaging approach).
The exponential terms in (\ref{CG})
could in principle give problems here so we consider this in the spirit of
\cite{ca,cb,gb}. Thus consider $\psi_{-} = \exp (-ikx + \phi)$ with
$\phi(x) = \int^x_{-\infty}\phi'(\xi)d\xi$. Now $\psi_{-}$ is going to have
the form $\psi_{-}= \exp(-S(X,k)/ \epsilon)$ and we can legitimately expect
$-S/{\epsilon} = -ikX/\epsilon + \phi(X/{\epsilon},k)$. For this to make
sense let us write (note $\epsilon x = X,\,\,\epsilon \xi = \Xi$) \be
\phi(X/{\epsilon},k) = \int_{-\infty}^{X/{\epsilon}} \phi'(\xi)d\xi =
\frac{1}{\epsilon}\int_{-\infty}^X\phi'(\Xi/\epsilon) d\Xi =
\frac{1}{\epsilon}(ikX - S(X,k)) \ . \label{CH} \ee Thus for
$\phi(X/\epsilon,k)$ to equal
${\epsilon}^{-1}(ikX - S(X,k))$ we must have \be
\int^X_{-\infty}\phi'(\Xi/\epsilon)d\Xi = ikX - S(X,k) \ , \label{CI} \ee
where $\phi'=\partial\phi/\partial x$. This says that $\phi'(\Xi/\epsilon)
= f(\Xi/\epsilon) = \tilde{f} (\Xi) + O(\epsilon)$ which is reasonable in
the same spirit that $u_n(T_j/\epsilon) = U_n(T_j) + O(\epsilon)$ was
reasonable before. Thus we have
\be
P = {\partial S \over \partial X} = ik - \phi' = v; \ \ \, P^2 - U = -k^2 \
, \label{CJ}
\ee
which corresponds to a dispersionless form of (\ref{CB}) in new variables.
Note here that $\phi'' = \epsilon \partial \phi'/\partial X \to 0$ as
$\epsilon \to 0$. Thus we have:
\begin{Proposition}
The manipulation of variables
$x,\,X$ in (\ref{CH}) and (\ref{CI}) is consistent and mandatory. It shows
that scattering information (expressed via $\phi$ for example) is related
to the dispersionless quantity $S$ or $P$. \end{Proposition}
Further we have isolated D-bar data ($\bar{\partial}_kS$) for S since for
$P = \sqrt{U - k^2}$ one can write for arbitrary $X_0$ (cf. (\ref{CI})) \be
ikX - S(X,k) = \int^X_{-\infty}\left(ik - P(\Xi,k)\right) \ d\Xi \ .
\label{CK} \ee
Replace $-\infty$ by $X_0$ for large negative $X_0$ to get then $ S =
\int^X_{X_0} Pd\Xi + ikX_0$.
Thus one expects a pole for $|k|\to\infty$ plus D-bar data along
$(-\sqrt{U},\sqrt{U})$. This is in the spirit of \cite{gb,gf} where it is
phrased differently. Observe that the expression $P = \sqrt{U - k^2}$
places us on a 2 sheeted Riemann surface with a cut along
$(-\sqrt{U},\sqrt{U})$ in the $k$-plane. Going around say $\sqrt{U}$ on the
+ sheet one has an opposite sides of the cut $P_{+} = i\sqrt{k^2 - U}$ and
$P_{-} = - i \sqrt{k^2 - U}$, so $P_{+} - P_{-} = \Delta P = 2 i\sqrt{k^2 -
U}$ which corresponds to D-bar data. This then leads to \be \Delta S =
S_{+} - S_{-} = \int^X_{X_0} \Delta P d\Xi = 2\int^X_{X_0}\sqrt{U(\Xi)-k^2}
\ d\Xi = 2Re S_{+} \ . \label{CL} \ee Note here that $P_{\pm}$ is real on
the cut with $P_{-} = -P_{+}$ (cf. Remark 3 for some general comments).
\\[2mm]\indent Consider next the situation of \cite{ca,gf,kc} with a
reduction of dKP to \be \Lambda:=\lambda^N = P^N + a_0P^{N-2} + ... +
a_{N-2} \ , \label{CM} \ee which is called dNKdV reduction.
Assume $\Lambda(P)$ has $N$ distinct real zeros $P_1 > P_2 > ... > P_N$
with $N-1$ interwoven turning points $\Lambda_k = \Lambda(\tilde{P}_k),
\,\,
P_{k+1} > \tilde{P}_k > P_k$. Assume,
as $X\to -\infty$, these Riemann invariants $\Lambda_k\to 0$ monotonically
so $\Lambda\to P^N$ with $(P - \lambda)\to 0$. Such situations are
considered in \cite{gf} and techniques from \cite{kc} are adapted to
produce formulas of the type ($W_m = \partial_P{\cal B}_m$)
\begin{eqnarray}
\label{CN1}
\nonumber
& &S(X,\mu) - P(X,\mu)X - {\cal B}_m(P(X,\mu),X)T_m \\ & &= -{1 \over 2\pi
i}\oint_{\Gamma}\frac{S(X,\lambda)
\partial_{\lambda}P(X,\lambda)}{P(X,\lambda) - P(X,\mu)}d\lambda \ ,
\end{eqnarray}
\begin{eqnarray}
\label{CN}
\nonumber
& & \frac{\partial S}{\partial \mu} - \frac{\partial P}
{\partial\mu}\left(X + W_m(P,X)T_m\right) = -{1 \over 2\pi i}\frac{\partial
P}
{\partial\mu}\oint_{\Gamma}\frac{S(X,\lambda)\partial_{\lambda}P(X,\lambda)}
{(P(X,\lambda)-P(X,\mu))^2} d\lambda \\
& &= -{\partial_{\mu}P \over 2\pi i}\oint_{\Gamma}
\frac{\partial_{\lambda}S(X,\lambda)}
{P(X,\lambda) - P(X,\mu)}d\lambda = \frac{\partial_{\mu}P}{\pi}
\int_{\Gamma_{+}}\frac{\partial_{\lambda}Im S(X,\lambda)}{P(X,\lambda) -
P(X,\mu)}d\lambda \ .
\end{eqnarray}
We see here the emergence of the Cauchy type kernel from (\ref{BF}) in an
important role. At $\lambda_k:=\lambda(\tilde P_k)$ where $S_{\mu}$ is
bounded and
$P_{\mu}\to \infty$ (\ref{CN}) yields Tsarev type generalized hodograph
formulas (cf. \cite{gf}). Thus such a formula is \be X + W_m( \tilde
P_k(X),X)T_m = -\frac{1}{\pi}\int_{\Gamma_{+}} \frac{\partial_{\lambda} Im
S(X, \lambda)}{P(X,\lambda) - \tilde P_k(X)}d\lambda \ . \label{YE} \ee
Here one has Riemann invariants
$\lambda_k=\lambda(\tilde{P}_k)$ where $\partial_P\Lambda = 0$ and there is
a collection $L$ of finite cuts through the origin of angles $k\pi/N\,\,
(1\leq k\leq N-1)$ in the $\lambda$ plane with branch points $\lambda_k$.
One takes $\Gamma$ to be a
contour encircling the cuts clockwise (not containing $\mu$) and sets
$\Gamma = \Gamma_{-} - \Gamma_{+}$ where + refers to the upper half plane.
It turns out that $S|_{\Gamma_{+}} = \bar{S}|_{\Gamma_{-}}$ and the
contours can be collapsed onto the cuts to yield the last integral in
(\ref{CN}) (see \cite{cl,gf} for pictures and details). By reorganizing the
terms in the integrals one can express now integrals such as (\ref{CN}) in
terms of D-bar data of $P$ on the cuts (cf. \cite{cl} for details). Thus
$P$ and $S$ are analytic in $\lambda$ for finite $\lambda$ except on the
cuts $L$ where there is a jump discontinuity $\Delta P$ (yielding $\Delta
S$ by integration in $X$ as in (\ref{CL})). We have seen for dKdV that
$\Delta P=2\sqrt{U-k^2}$ on $(-\sqrt{U}, \sqrt{U})$ and other dNKdV
situations are similar (cf. \cite{cl,gf}). Moreover there is no need to
restrict ourself to one time variable $T_m$ in (\ref{CN}), or for that
matter to dNKdV. Indeed the techniques, leading to formulas of the type
(\ref{CN}) are based on \cite{kc} (cf. also \cite{ca}), and apply equally
well to dKP provided the D-bar data $\bar{\partial}P$ for dKP lie in a
bounded set $\Omega$ (cf. \cite{cl} and remarks below). In this respect,
concerning dKP, one notes first that the transformation $t\to -t,\,\,U\to
-U$ sends dKP-1 to dKP-2 so one expects that any D-bar data for $S$ or $P$
will be the same. Secondly, we know from \cite{ca,kc} that for large
$|\lambda|$ there will be formulas of the type (\ref{BF}). Let us assume
that $\bar{\partial}P$ (and hence $\bar{\partial}S$) is nontrivial only for
a region $\Omega$ where say $|\lambda|\leq M < \infty$, and let $\Gamma$ be
a contour enclosing $|\lambda|
\leq M$. Then without regard for the nature of such data (Riemann- Hilbert
data, poles, simple nonanalyticity, etc.) one can in fact derive formulas
(\ref{CN}) following \cite{ca,gf,kc} (see \cite{cl} for details). We cannot
collapse on cuts as in the last equation of (\ref{CN}) but we can think of
the other formulas in (\ref{CN}) as integrals over D-bar data
$\bar{\partial}P$ or $\bar{\partial}S$. It remains open to describe D-bar
data for the dKP situations however. The idea of some kind of limit of
spiked cut collections arises but we have not investigated this. In summary
(cf. \cite{cl} for more detail) we can state: \begin{Proposition} For dNKdV
(or dKP with given
bounded D-bar data) one can write
\begin{eqnarray}
S(X,\mu) = P(\mu)X + \sum_2^{\infty}{\cal B}_n(P(\mu),X)T_n - \frac{1}{2\pi
i}\oint_{\Gamma}
\frac{S(X,\lambda)\partial_{\lambda}P(X, \lambda)} {P(X,\lambda) -
P(X,\mu)}d\lambda \ ,
\label{ZD}
\end{eqnarray}
where $\Gamma$ encircles the cuts (or D-bar data) clockwise. The
determination of D-bar data here is to be made via analysis of the
polynomial $\lambda^N$ in (\ref{CM}) for dNKdV and in this situation one
can again
collapse (\ref{CN}) to the cuts.
\end{Proposition}
{\bf Remark 3} $ \, \, $ Now perhaps the main point of this has been to
show that $S$ can be characterized via D-bar data of $S$ or $P$. In general
we do not expect D-bar or scattering data for NKdV or KP to give D-bar data
for $S$. The case of KdV is probably exceptional here and the situation of
\cite{gb,za} where spectral data for a system of nonlinear Schr\"odinger
equations is related to the Benney or dKP hierarchy involves a different
situation (the spectral data is not related to KP). Further since KP-1 and
KP-2 have vastly different spectral or D-bar properties, and both pass to
the same dKP, one does not expect the spectral data to play a role. We note
also that spectral data is created by the potentials via asymptotic
conditions
for example (and vice versa) whereas D-bar data for dNKdV, arising from the
polynomial $\lambda^N$, is a purely algebraic matter. \\[3mm]\indent There
is an interesting way in which D-bar data for $P$ or $S$ can be exploited.
Thus, given that $S$ has D-bar data as indicated above in a bounded region,
one can say that for $|\lambda|$ large and with some analytic function $A$
of $\lambda$,
\begin{eqnarray}
\nonumber
S &=& A + {1 \over 2\pi i}\int\int\frac{\bar{\partial}_{\zeta}S} {\zeta -
\lambda}d\zeta\wedge d\bar{\zeta} \\ &=& A - {1 \over 2\pi
i}\sum_0^{\infty}\frac{1}{\lambda^{j+1}}\int\int
\zeta^j\bar{\partial}_{\zeta}S \ d\zeta\wedge d\bar{\zeta} \ , \label{CO}
\end{eqnarray}
and from ({\ref{action}), we get
\be
\partial_j F = {j \over 2\pi i}\int\int\zeta^{j-1} \bar{\partial}_{\zeta}S
\ d\zeta\wedge d\bar{\zeta} \ . \label{CP} \ee This is very useful
information about $F_j$, since from the computation of (\ref{CP}) one could
in principle compute all of the functions $F_{ij}$ for example. In fact,
for dKdV, and possibly some other dNKdV situations, one can obtain a direct
formula for the $F_{1j}$, which we know to generate all the $F_{ij}$ via
Lemma 1. Thus for dKdV we know from (\ref{CL}) that $\Delta S =
\int_{X_0}^X\Delta P dX' = 2\int^X_ {X_0}\sqrt{U(X',T)-k^2}dX'$ on the cut
$L = (-\sqrt{U},\sqrt{U})$ in the k-plane, and $\bar{\partial}S=\Delta S
(i/2)\delta_L$ where $\delta_L$ is a suitable delta function on $L$.
Adjusting variables ($\zeta^2 = -k^2$ etc.) one can write then \be
\partial F_j = -\frac{j(i)^{j-1}}{\pi}\partial\int_{-\sqrt{U}}^{\sqrt{U}}
k^{j-1}\int_{X_0}^X\sqrt{U(X',T) - k^2} \ dX'dk \ . \label{CQ} \ee The only
$X$ dependence here is visible and one can differentiate under the integral
sign in (\ref{CQ}). Since all $F_{1 \ 2m} =0$ by the residue formula
(\ref{AZ}), for example we take $j = 2n-1$ to obtain \begin{eqnarray}
\nonumber
& & F_{1 \ 2n-1} =
(-1)^n\frac{2n-1}{\pi}\int_{-\sqrt{U}}^{\sqrt{U}}k^{2n-2} \sqrt{U-k^2}dk =
\\ & &=(-1)^n\frac{(2n-1)U^n}{\pi}\int_{-\frac{\pi}{2}}^
{\frac{\pi}{2}}\sin^{2n-2}\theta \cos^2\theta d\theta = (-1)^n
\left(\frac{U}{2}\right)^n\prod_1^n\frac{2l-1}{l} \label{CR} \end{eqnarray}
(here $k = a\sin\theta$ and $a^2 = U$).
We summarize this in (cf. \cite{cl} for more detail) \begin{Theorem} For
any situation with
bounded D-bar data for $S$ one can compute $F_j$ from (\ref{CP}). In the
case of dKdV with $\Delta S$ as indicated one can use (\ref{CR}) to
determine directly the $F_{1 \, 2n-1}$ which will generate all the
functions $F_{ij}$.
Generally, if $\bar{\partial}S = \int_{X_0}^X\bar{\partial}P(X',T,
\zeta,\bar{\zeta})dX'$ (as in (\ref{CQ})), then from (\ref{CP}) $F_{1j} =
(j/2\pi i)\int\int\zeta^{j-1}\bar{\partial}Pd\zeta\wedge d\bar{\zeta}$, and
formally, $\partial_n\partial_j F = (j/2\pi i)
\int\int\zeta^{j-1}\bar{\partial}{\cal B}_n d\zeta\wedge d\bar{\zeta}$
since formally $\partial_n\bar{\partial}S = \bar{\partial}\partial_n S =
\bar{\partial}{\cal B}_n$.
\end{Theorem}
{\bf Remark 4} $\,\,$ The calculations using (\ref{CQ}) in fact agree with
the determination of $F_{1 \ 2n-1}$ from residue calculations as in Lemma 1
(note here $2U_2= -U$ when adjusting notations between Sections 2 and 4).
In general one may have some difficulties in determining
$\bar{\partial}{\cal B}_n$ or $\bar{\partial}P$ for example, so the
formulas (\ref{CQ}) and (\ref{CR}) may be exceptional in this regard.
However for dKdV
we have checked the validity of these formal integrals for $F_{nj}$
involving $\bar{\partial}{\cal B}_n$ for a number of terms (e.g.
$F_{31},\,\,F_{51},\,\,F_{33},\,\,F_{35}$, and $F_{53}$) against the
results of residue calculations. Here one can write for $n$ odd
$\bar{\partial}{\cal B}_n = \bar{\partial}\zeta^n_{+} = \Delta\zeta^n_{+} =
P^{-1}\zeta^n_{+}\Delta P$ and one gets for $n = 2k-1,\,\,j = 2m-1$ \be
F_{nj} = (-1)^m\frac{2(2m-1)U^{k+m-1}}{\pi}\int_0^{\frac{\pi}{2}}
\sin^{2m-2}\theta\hat{\zeta}^{2k-1}_{+}(\cos\theta)\cos\theta d\theta \ ,
\label{NN}
\ee
where $a^{2k-1}\hat{\zeta}^{2k-1}_{+}(\cos\theta)=\zeta^{2k-1}_{+}(P)$ for
$P = a\cos\theta\,\,(a^2 = U)$.
Let us mention also that one can easily write down the tables of $F_{ij}$
for dNKdV from the table (\ref{AX}). Thus for dNKdV we have a reduction of
the table (\ref{AX}) given via $F_{Nm}=F_{mN}=0$ for all $m$.

\par\medskip\medskip

\noindent
{\bf Acknowledgment}
We thank Mr. Seung Hwan Son for his Mathematica programs yielding
(\ref{AX}) from (\ref{AV}) and $F_{1 \ 2n-1}$ for the dKdV from (\ref{AZ}).
The work of YK is partially supported by an NSF grant DMS-9403597.


\begin{thebibliography}{99}

\bibitem{ab} M. Ablowitz and P. Clarkson, Solitons, nonlinear evolution
equations and inverse scattering, Cambridge Univ. Press, 1991.
%
\bibitem{aa} M. Adler and P. vanMoerbeke, Comm. Math. Phys., 147 (1992), 25-56.
%
\bibitem{ac} S. Aoyama and Y. Kodama,
Physics Lett. 295B, (1992), 190-198; Phys. Lett. 278B (1992), 56-62; Mod.
Phys. Lett. 9A (1994), 2481-2492.
%
\bibitem{af} S. Aoyama and Y. Kodama,
Topological Landau-Ginzburg theory with a rational potential and the
dispersionless KP hierarchy, hep-th/9505122. %
\bibitem{bd} M. Boiti, L. Martina, and F. Pempinelli, Multidimensional
localized solitons, preprint, 1993. %
\bibitem{ca} R. Carroll,
Jour. Nonlinear Sci., 4 (1994), 519-544; Teor. i Matem. Fizika, 99 (1994),
220-225.
%
\bibitem{cb} R. Carroll,
On dispersionless Hirota type equations, hep-th 9410063, Proc. NEEDS 94,
World Scientific, to appear. %
\bibitem{ce} R. Carroll,
Topics in soliton theory, North-Holland, 1991. %
\bibitem{cf} R. Carroll and B. Konopelchenko, Lett. Math. Phys., 28 (1993),
307-319.
%
\bibitem{cj} R. Carroll,
Some connections between KdV and physics, in preparation. %
\bibitem{cl} R. Carroll and S. Son,
Remarks on dispersionless D-bar, in preparation. %
\bibitem{dp} B. Dubrovin,
Nucl. Phys. B, 374 (1992), 627-689; Comm. Math. Phys., 145 (1992), 195-207;
Integrable systems, Birkhauser, 1993, pp. 313-359; hep-th 9303152;
SISSA-89/94/FM, hep-th/9407018; Integrable quantum field theories, Plenum,
1993, pp. 283-302.
%
\bibitem{fb} A. Fokas and P. Santini,
Physica 44D (1990), 99-130.
%
\bibitem{gb} V. Geogdzhaev,
Teor. i Matem. Fizika, 73 (1987), 255-263; Singular limits of dispersive
waves, Plenum,
1994, pp. 53-59.
%
\bibitem{gf} J. Gibbons and Y. Kodama,
Singular limits of dispersive waves, Plenum, 1994, pp.61-66. %
\bibitem{gh} J. Gibbons,
Dynamical Problems in Soliton Systems, Springer, 1985, pp. 36-41. %
\bibitem{ka} Y. Kodama,
Phys. Lett. 129A (1988), 223-226 ; Prog. Theor. Phys., Supp. 94 (1988),
184-194.
%
\bibitem{kc} Y. Kodama and J. Gibbons,
Phys. Lett. 135A (1989), 167-170; Proceedings IV Workshop on Nonlinear and
Turbulent Processes in Physics, World Scientific, 1990, pp. 166-180. %
\bibitem{kd} Y. Kodama,
Lecture Lyon Workshop, July, 1991, unpublished. %
\bibitem{kg} B. Konopelchenko,
Solitons in multidimensions, World Scientific, 1993; Introduction to
multidimensional integrable systems, Plenum,1992. %
\bibitem{kk} I. Krichever,
Comm. Math. Phys., 143 (1992), 415-429; Comm. Pure Appl. Math., 47 (1994),
437-475; New symmetry principles in quantum field theory, Plenum, 1992, pp.
309-327.
%
\bibitem{lb} J. Leon,
Jour. Math. Phys., 35 (1994), 3504-3524. %
\bibitem{ma} S. Manakov,
Physica 3D (1980), 149-157.
%
\bibitem{tb} K. Takasaki and T. Takebe,
Inter. Jour. Mod. Phys. A, Supp., 1992, pp. 889-922. %
\bibitem{tc} K. Takasaki and T. Takebe,
hep-th 9405096.
%
\bibitem{za} V. Zakharov,
Funct. Anal. i Prilozh., 14 (1980), 15-24.

\end{thebibliography}
\end{document}